# Modulating Anisotropic Magnetism of Layered $CuCrP_2S_6$ Single Crystal via Selenium Substitution

I. S. Eid[1,2] and Li Song[1,*]

[1]National Synchrotron Radiation Laboratory, University of Science and Technology of China, Hefei, Anhui 230029, China

[2]Department of Physics, Faculty of Science, Benha University, Benha 13518, Egypt

*E-mail: song2012@ustc.edu.cn

**ABSTRACT**

As one of typical layered structures, antiferromagnetic $CuCrP_2S_6$ single crystal has high potential for magnetoelectric devices and spintronic multi-terminal chips due to its unique magnetic anisotropy. However, to tune the anisotropic magnetism is still challenge. Herein, high quality single crystals of $CuCrP_2(S_{1-X}Se_X)_6$ (where X=0, 0.02, 0.05, 0.1, and 0.2) were grown via chemical vapor transport route, and the effect of selenium substitutions on the magnetic anisotropies in $CuCrP_2S_6$ were systematically investigated. The structural characterizations revealed that the $CuCrP_2(S_{1-X}Se_X)_6$ single crystals could remain the pristine monoclinic crystal phase with Se uniform dispersion in whole $CuCrP_2S_6$. With Se substitution, the oxide states of P and S elements exhibited increased. The magnetic anisotropic results measured in normal and parallel directions between the crystallographic plane (**ab**) and the applied magnetic field (**H**) showed that the Néel temperatures ($T_N$) and temperatures at maximum magnetic susceptibilities ($T_{Max}$) in **H⊥ab** and **H∥ab** decreased with increasing the selenium doping amounts in $CuCrP_2(S_{1-X}Se_X)_6$ single crystals. Notably, their anisotropic behavior in paramagnetic phase tends to be isotropic with increasing selenium concentrations. Moreover, the transitions of the spin flop, which appear only in case of **H⊥ab**, was greatly suppressed with increasing the selenium concentrations in $CuCrP_2(S_{1-X}Se_X)_6$ until disappearing at X=0.1 and 0.2. The spin polarized density functional theory calculations further confirmed that the asymmetry between spin up and spin down states in $CuCrP_2S_6$ could be highly modulated via Se substitution, resulting in the tunable anisotropic magnetism of hybridized $CuCrP_2(S_{1-X}Se_X)_6$. Our findings make these materials to be more interesting in spintronic applications.

## I. INTRODUCTION

Since the successful fabrication of a graphene monolayer, two-dimensional layered materials have occupied a remarkable attention due to their unique electronic, optical, and magnetic properties [1, 2]. Among them, Transition Metal Phosphorous Trichalcogenides ($M_2P_2X_6$; X=S and Se) with untraditional magnetic and electronic properties are more promising for many specific applications [3-9]. Generally, $M_2P_2X_6$ compounds are structurally composed of van der Waals alternating stacking of a few or several layers dependent of their thickness [10, 11]. In principle, such $M_2P_2X_6$ can be geometrically formed in a monoclinic crystal structure with various magnetic properties based on the electronic occupations in 3d of transition metals [12].

Recently, $CuCrP_2S_6$ as one of typical $M_2P_2X_6$ materials has attracted many efforts because of the presence of both the intrinsic antiferromagnetic and antiferroelectric features in its structure [13-15]. In principle, $CuCrP_2S_6$ is crystallized in a monoclinic unit cell with alternating periodic arrangements of Cu and Cr atoms through its honeycomb sublattice, in which the antiferroelectricity generated by Cu and the antiferromagnetism generated by Cr [16, 17]. Experimentally, Colombet et al. [18] found that $CuCrP_2S_6$ behaved as an antiferromagnetic material below Néel temperature ($T_N$)=30 K. Selter et al. [19] further reported that the four stacking layers of $CuCrP_2S_6$ showed the antiferromagnetic characteristics below similar temperature as well as a positive value of Curie-Weiss temperature. More interestingly, $CuCrP_2S_6$ can also exhibit magnetic anisotropies in its intrinsic antiferromagnetic phase. Wang et al. [20] found that the antiferromagnetism in $CuCrP_2S_6$ was displayed below $T_N$=32 K with considerable magnetic anisotropies through the crystallographic directions. Additionally, they reported the coexistence of spin flop transitions which illustrated a high anisotropy with changing the crystallographic directions.

In fact, such intrinsic magnetic properties of $M_2P_2X_6$ materials can be tuned by applying an external pressure or chemical substitutions [21, 22], subsequently exhibiting various anisotropies [23, 24]. For example, parent systems of $Ni_2P_2S_6$ and $Fe_2P_2S_6$ showed different magnetic anisotropies in comparison with $(Fe_{1-X}Ni_X)_2P_2S_6$ as an intermediate system between them [25]. Besides, the effect of the chemical substitutions between Ni and Mn ions was studied on the magnetic anisotropic characteristics in $(Mn_{1–X}Ni_X)_2P_2S_6$ layered single crystals in comparison with their parent counterparts ($Mn_2P_2S_6$ and $Ni_2P_2S_6$) [21, 26]. Abraham *et al.* [27] investigated the magnetic anisotropy in mixed transition metallic cations system of $MnNiP_2S_6$ which was compared with $Mn_2P_2S_6$. Moreover, $CuCr_{1-X}In_XP_2S_6$ compound elucidated a modulation in its magnetic anisotropy via the substitution between Cr and In ions [28]. Recently, some evolutions of magnetic anisotropies in different $M_2P_2X_6$ systems have been investigated by the means of elemental substitutions [29, 30]. Therefore, it is highly desirable to develop a controllable approach to modulate the magnetism of $M_2P_2X_6$ for more applications.

Herein, we demonstrate a chemical vapor transport route to facile modulate the anisotropic magnetism of $CuCrP_2S_6$ via in-situ selenium substitution. Scanning electron microscopy (SEM) and energy dispersive X-ray spectroscopy (EDS) were used for morphological and compositional characterizations of the grown $CuCrP_2(S_{1-X}Se_X)_6$ single crystals. X-ray diffraction (XRD), high resolution transmission electron microscope (HRTEM), selected area electron diffraction (SAED) were employed to characterize their single crystallinity. The chemical bonding states in the crystals were studied by X-ray photoelectron spectroscopy (XPS). The experimental magnetic measurements revealed that the anisotropic antiferromagnetism of $CuCrP_2(S_{1-X}Se_X)_6$ can highly modulated by Se doping amounts. Moreover, spin polarized density functional theory (DFT) calculations have been performed to study the effect of Se substitutions on the magnetic properties of the parent compound $CuCrP_2S_6$.

## II. RESEARCH METHODOLOGY

### 1. EXPERIMENTAL METHODOLOGY

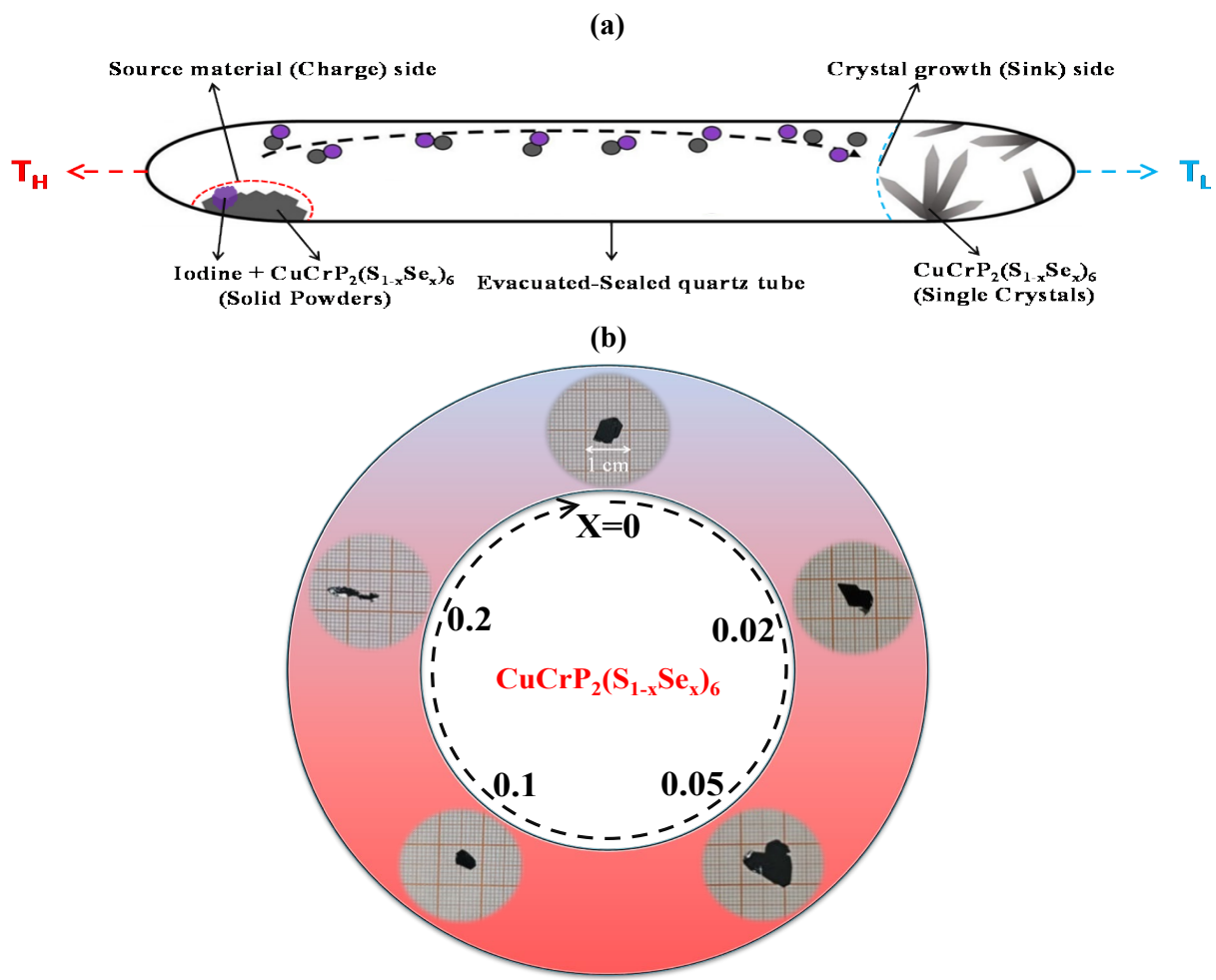

**Figure 1: (a)** Schematic illustration of the tube during the CVT reaction for synthesizing $CuCrP_2(S_{1-X}Se_X)_6$ single crystals, and **(b)** Optical images of the grown $CuCrP_2(S_{1-X}Se_X)_6$ crystals for (X=0, 0.02, 0.05, 0.1, and 0.2).

High qualities of $CuCrP_2(S_{1-X}Se_X)_6$ single crystals (X=0, 0.02, 0.05, 0.1, and 0.2) were grown by chemical vapor transport (CVT) technique. The solid powders of the elements in the stoichiometric molar proportions of Cu, Cr, P, S, and Se were mixed together as a source material with iodine (~ 30-31 mg) as a transport agent and were sealed in a 15 cm long evacuated sealed quartz tube with 1cm inner diameter as shown in **Figure 1(a)**. This tube was transported horizontally into a two-zone furnace in which the two zones were defined as a source (charge) side with a higher temperature ($T_H$) and sink (crystal growth) side with a lower temperature ($T_L$). The controlled temperature profile of the furnace was set up where the charge and sink sides were raised within 360 min from room temperature to 300 °C for satisfying the pre-reaction of P, S, and Se elements with the transition elements [25]. Then, after 1 day, the charge side temperature was set up at 690 ºC while the sink side became at 620 ºC. Finally, both sides were dewelled at these temperatures for 6 days (crystal growth phase) and then, the furnace was slowly cooled to the room temperature. The optical images of the grown crystals extracted from the tube, are demonstrated as shown in **Figure 1(b)**.

## 2. CALCULATION METHODS

DFT calculations with spin polarization effects were employed by using Vienna ab initio simulation package (VASP) [31]. The PBE functional was used to express the ion-electron interaction and generalized gradient approximation (GGA) [32] which were characterized by projector augmented wave (PAW) method [33, 34]. The energy cut off of a plane wave was set to 385 eV. Also, Gaussian smearing was set to equal 0.05 eV. The energy convergence was considered in precision of $10^{-6}$ eV. All atoms were fully relaxed in the structure until the force became less than 0.01 eV/Å. In addition, the Brillouin zone was performed with a 6 x 3 x 3 Monkhorst-Pack of k-points scheme [35] for the geometrical optimization of the structures while it became 12 x 6 x 6 k-points for density of state (DOS) calculations.

## III. RESULTS AND DISCUSSION

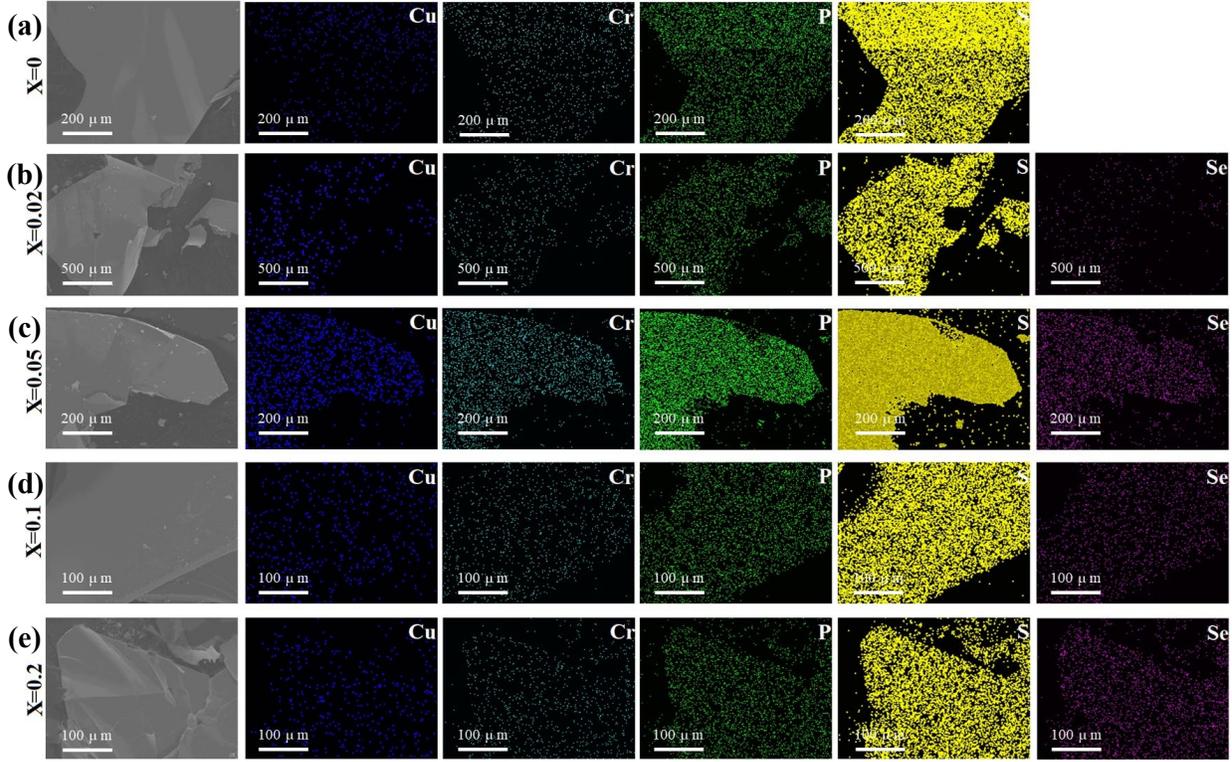

**Figure 2:** SEM images and their corresponding EDS elemental mappings of the cleaved grown CuCrP$_2$(S$_{1-X}$Se$_X$)$_6$ crystals for (X=0 **(a)**, 0.02 **(b)**, 0.05 **(c)**, 0.1 **(d)**, and 0.2 **(e)**).

The layered morphological nature of all cleaved grown CuCrP$_2$(S$_{1-X}$Se$_X$)$_6$ crystals for (X=0 **(a)**, 0.02 **(b)**, 0.05 **(c)**, 0.1 **(d)**, and 0.2 **(e)**) is demonstrated according to SEM images, as shown in **Figure 2**. Their corresponding EDS measurements illustrate a homogenous distribution of all elements through the whole crystalline surface. Besides, the compositional analysis of the atomic percentages of all elements inside the crystalline surfaces is recorded, as seen in **Table 1**. It is observed that there is an acceptable deviation in the atomic percentage values of the elements between the experimentally used precursor values and measured EDS values. This is due to several parameters associated with CVT reactions, such as the duration and the temperature of the crystal growth, the transport agent type and the temperature gradient between the source and sink sides [36].

**Table 1**: Atomic percentages of elements in CuCrP$_2$(S$_{1-X}$Se$_X$)$_6$ crystals.

| X$_{Se}$ in CuCrP$_2$(S$_{1-X}$Se$_X$)$_6$ | Cu (At %) | Cr (At %) | P (At %) | S (At %) | Se (At %) |
|---|---|---|---|---|---|
| 0 | 11.25 | 11.81 | 18.09 | 58.86 | 0 |
| 0.02 | 11.13 | 12.77 | 18.23 | 56.97 | 0.9 |
| 0.05 | 11.84 | 11.31 | 18.85 | 55.60 | 2.41 |
| 0.1 | 12.83 | 12.24 | 17.60 | 52.91 | 4.43 |
| 0.2 | 11.48 | 11.77 | 18.20 | 50.42 | 8.13 |

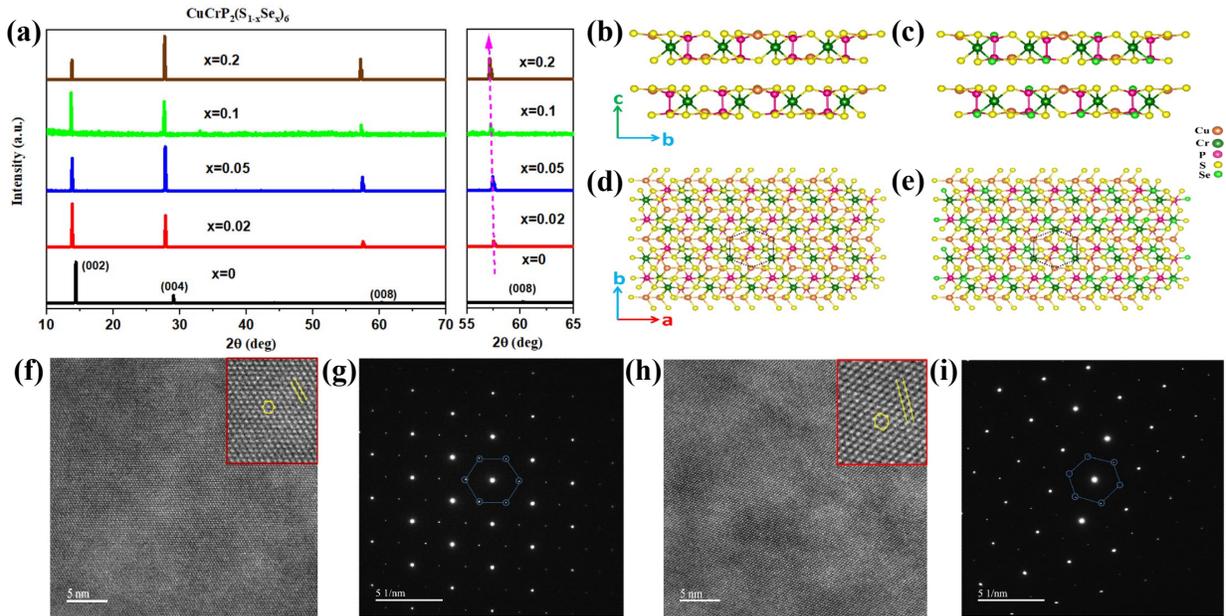

**Figure 3:** Crystal Structural Characterization. **(a)** XRD pattern of CuCrP$_2$ (S$_{1-X}$Se$_X$)$_6$ single crystals (X=0, 0.02, 0.05, 0.1, 0.2). Atomic structure models of CuCrP$_2$S$_6$ (X=0) and CuCrP$_2$S$_{4.8}$Se$_{1.2}$ (X=0.2) from side view **(b, c)**, as well as top view **(d, e)**. HRTEM image of CuCrP$_2$S$_6$ (X=0) **(f)** and CuCrP$_2$S$_{4.8}$Se$_{1.2}$ (X=0.2) **(h)**, and their corresponding SAED patterns **(g)** and **(i)**.

According to XRD patterns of CuCrP$_2$(S$_{1-X}$Se$_X$)$_6$ single crystals as shown in **Figure 3(a)**, the diffraction peaks (002), (004), and (008) of pristine CuCrP$_2$S$_6$ have been indexed well in agreement with the literatures [15, 20, 37]. It can be observed that the undoped CuCrP$_2$S$_6$ has a monoclinic crystal phase with *Pc* space group [20]. In addition, it is clearly seen that these peaks have been shifted to lower Bragg's angles under selenium doping because the ionic radius of sulfur is smaller than that of selenium. Also, with increasing the Se substitution from X = 0.02 to X = 0.2, there is a monotonic gradual observable deviation of (008) peaks towards the lower Bragg's angles in agreement with literatures [29, 30]. The peaks shift to the lower Bragg's angles under Se doping indicates to the elongation occurred along (c) crystal axis due to the difference in ionic radii of S and Se. **Figure 3 ((b)-(e))** represent the crystal structure visualizations of undoped CuCrP$_2$S$_6$ (side view **(b)** and top view **(d)**) and Se doped CuCrP$_2$S$_6$ for X=0.2 (side view **(c)** and top view **(e)**). In the atomic models, a regular alternating arrangement of Cu and Cr cations in hexagonal rings is demonstrated through the whole structure for undoped and Se doped CuCrP$_2$S$_6$ crystals. In **Figure 3**, HRTEM images and their corresponding SAED patterns of pristine CuCrP$_2$S$_6$ (**(f)** and **(g)**) and Se doped CuCrP$_2$S$_6$ at X=0.2 (**(h)** and **(i)**) elucidate the layered morphological nature and the uniform atomic distribution with hexagonal shaped rings throughout the crystal structure without any clustering of Se dopant in agreement with the above SEM obversions.

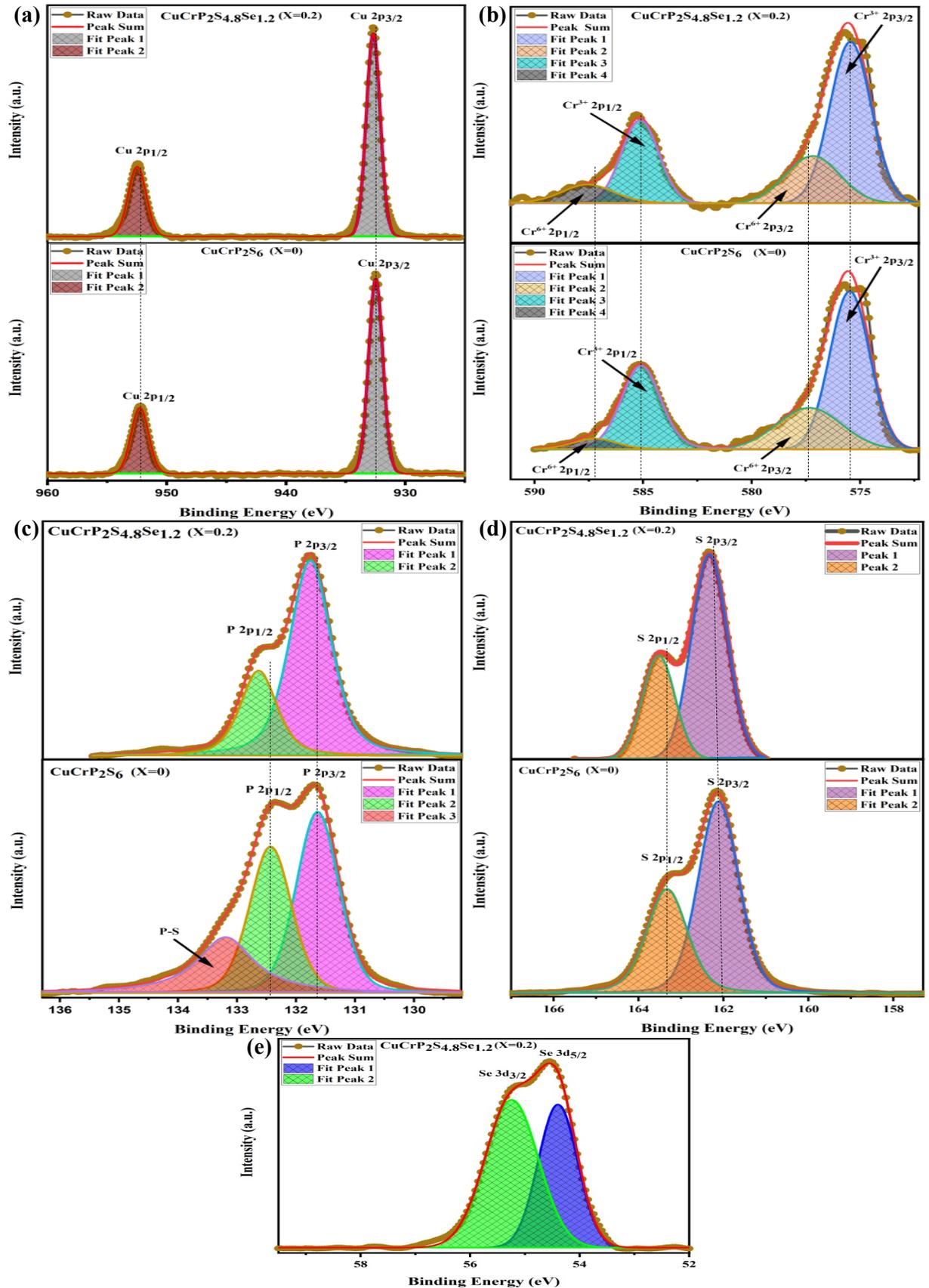

**Figure 4**: XPS spectra of **(a)** Cu 2p, **(b)** Cr 2p, **(c)** P 2p, **(d)** S 2p and **(e)** Se 3d of pure $CuCrP_2S_6$ and Se doped $CuCrP_2S_6$ ($CuCrP_2S_{4.8}Se_{1.2}$).

For optimizing the evolution of chemical bonding states under the Se doping effect in $CuCrP_2S_6$, we employed XPS for characterizing $CuCrP_2(S_{1-x}Se_x)_6$ crystals for X=0 (pristine $CuCrP_2S_6$) and X=0.2 ($CuCrP_2S_{4.8}Se_{1.2}$). As shown in **Figure 4(a)**, we have observed that the Cu 2p peaks of doped and undoped $CuCrP_2S_6$ include the same two peaks assigned to Cu $2p_{1/2}$ and Cu $2p_{3/2}$. The Cu $2p_{1/2}$ peak is detected at 952.27 eV and 952.47 eV, while Cu $2p_{3/2}$ peak is positioned at 932.49 eV and 932.68 eV for pristine $CuCrP_2S_6$ and Se doped $CuCrP_2S_6$, respectively. This indicates that Cu 2p peaks for doped $CuCrP_2S_6$ are slightly shifted to the higher binding energies compared to the same peaks in undoped $CuCrP_2S_6$. Also, **Figure 4(b)** shows that in doped and undoped $CuCrP_2S_6$, Cr 2p exhibited similar four peaks assigned as $Cr^{3+}$ $2p_{1/2}$, $Cr^{3+}$ $2p_{3/2}$, $Cr^{6+}$ $2p_{1/2}$, and $Cr^{6+}$ $2p_{3/2}$. For pristine $CuCrP_2S_6$, $Cr^{3+}$ $2p_{1/2}$ and $Cr^{3+}$ $2p_{3/2}$ are located at 585.10 eV and 575.48 eV, respectively. The two peaks are found at 585.14 eV and 575.45 eV for $CuCrP_2S_{4.8}Se_{1.2}$. It is clearly seen that there is no a considerable shift for these peaks in both doped and undoped $CuCrP_2S_6$. In addition, $Cr^{6+}$ $2p_{1/2}$ peak is detected at 587.23 eV and 587.55 eV for undoped and doped $CuCrP_2S_6$, respectively, while $Cr^{6+}$ $2p_{3/2}$ peak is seen at 577.35 eV and 577.19 eV for undoped and doped $CuCrP_2S_6$, respectively. It is further noticed that $Cr^{6+}$ $2p_{1/2}$ peak in Se doped $CuCrP_2S_6$ is shifted to the higher binding energies compared to the pristine structure while $Cr^{6+}$ $2p_{3/2}$ peak in doped $CuCrP_2S_6$ is very slightly shifted to the lower binding energies compared to the pristine $CuCrP_2S_6$. Moreover, as illustrated in **Figure 4(c)**, the P 2p peaks of pristine $CuCrP_2S_6$ are involved into three peaks assigned as P-S, P $2p_{1/2}$, and P $2p_{3/2}$. These peaks are respectively detected at 133.20 eV, 132.43 eV, and 131.64 eV in undoped $CuCrP_2S_6$. In comparison with pristine $CuCrP_2S_6$, there are only two peaks of P 2p in doped $CuCrP_2S_6$ named as P $2p_{1/2}$ and P $2p_{3/2}$ positioned at 132.64 eV and 131.76 eV, respectively. Comparison of both structures shows that the extra peak in the pristine structure refers to the formation of the chemical bonds on its crystalline surface between phosphorus and sulfur. Besides, it is demonstrated that P $2p_{1/2}$ and P $2p_{3/2}$ peaks for doped $CuCrP_2S_6$ are shifted to the higher binding energies compared with the same peaks in undoped $CuCrP_2S_6$ hence indicating that Se substitution results in donation of electrons to the pristine system resulting in increase in oxidation state. Also, **Figure 4(d)** illustrates the two peaks of S 2p (S $2p_{1/2}$ and S $2p_{3/2}$) for undoped and doped $CuCrP_2S_6$. S $2p_{1/2}$ peak is detected at 163.33 eV and

163.52 eV while S $2p_{3/2}$ peak is localized at 162.12 eV and 162.33 eV for undoped and doped $CuCrP_2S_6$, respectively. It is shown that S $2p_{1/2}$ and S $2p_{3/2}$ peaks for doped $CuCrP_2S_6$ are shifted to higher binding energies compared with the similar peaks in the pristine $CuCrP_2S_6$ hence resulting in increase in oxidation state as well. Furthermore, as demonstrated in **Figure 4(e)**, Se 3d peaks in Se doped $CuCrP_2S_6$ include two peaks assigned as Se $3d_{3/2}$ and Se $3d_{5/2}$. Se $3d_{3/2}$ and Se $3d_{5/2}$ peaks are located at 55.24 eV and 54.40 eV. All above XPS peaks have been indexed in accordance with the previous literatures [23, 38-42]. The observed displacements of Cu, Cr, P, and S peaks to the higher binding energy after Se doping show that selenium atoms have a large tendency to loss more electrons leading to a charge transfer to its neighbors [43].

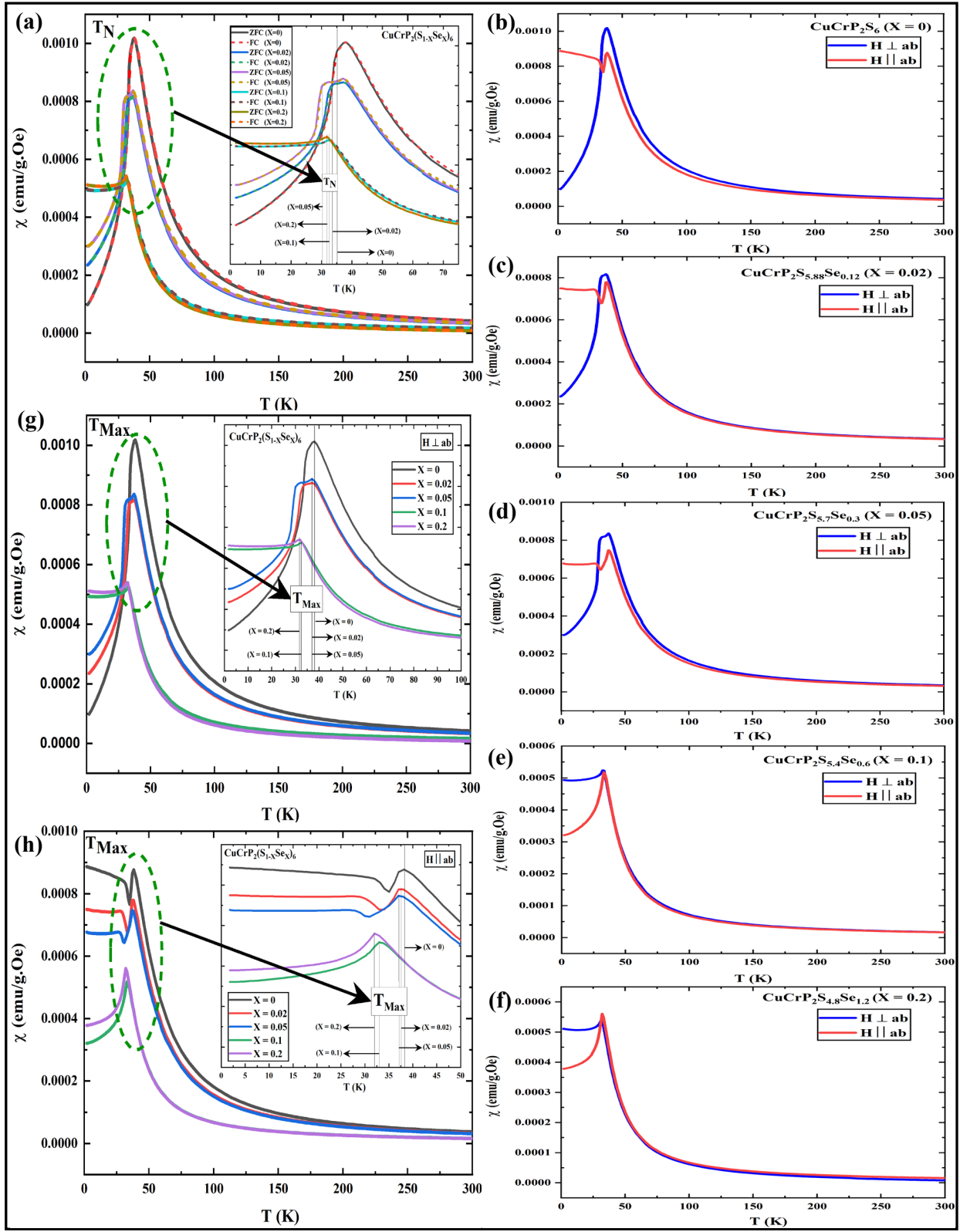

**Figure 5:** Magnetic susceptibility (χ(emu/g.Oe)) as a function of the temperature (T (K)) for an applied magnetic field of 1000 Oe **(a)** Zero Field Cooling (ZFC) and Non- Zero Field Cooling (FC) for CuCrP$_2$(S$_{1-X}$Se$_X$)$_6$ crystals. The inset in **(a)** represents a zoomed view of the Néel

temperature ($T_N$) positions. **(b)**, **(c)**, **(d)**, **(e)**, and **(f)** show the χ vs. T curves for $CuCrP_2(S_{1-X}Se_X)_6$ crystals at X=0, 0.02, 0.05, 0.1, and 0.2, respectively, in both normal and parallel directions of the applied magnetic field **H** to the **ab** crystal plane. **(g)** and **(h)** are χ vs. T curves in **H ⊥ ab** and **H ∥ ab**, respectively, for $CuCrP_2(S_{1-X}Se_X)_6$ crystals. The insets in **(g)** and **(h)** indicate to the positions of the temperature at maximum magnetic susceptibility ($T_{Max}$).

**Figure 5(a)** demonstrates the magnetic susceptibility (χ) as a function of temperature (T) in both Zero Field Cooling (ZFC) and Non-Zero Field Cooling (FC) plotted at 1000 Oe of applied magnetic field (**H**) to determine the Néel Temperature ($T_N$) of $CuCrP_2(S_{1-X}Se_X)_6$ crystals, where X=0, 0.02, 0.05, 0.1, and 0.2. It is clearly seen that with increasing the doping ratios of selenium in $CuCrP_2(S_{1-X}Se_X)_6$ structures, the values of $T_N$ decrease in agreement with the literature [30], except at X=0.05 which has deviated from this behavior as illustrated in **Table 2**. Also, it can be observed that the values of the magnetic susceptibility have still unchanged in both ZFC and FC measurements. In addition, from **Figure 5(b)** to **Figure 5(f)**, an external magnetic field (**H**) with intensity 1000 Oe was applied in the normal and parallel directions to the crystallographic plane **ab**, respectively, for studying isotropic and anisotropic magnetic properties of $CuCrP_2(S_{1-X}Se_X)_6$ crystals. Moreover, the behavior of temperature-dependence magnetic susceptibility relations in case of **H ⊥ ab** differs from that in case of **H ∥ ab** confirming the existence of anisotropic characteristics in $CuCrP_2(S_{1-X}Se_X)_6$ structures. Besides, with increasing the selenium doping amounts, the anisotropic characteristics in the paramagnetic phase ($T>T_N$) tend to be semi-isotropic unlike the antiferromagnetic phase ($T_N>T$) in which the isotropic characteristics remain dominant. Furthermore, the antiferromagnetic profile for **H ⊥ ab** is longer than that for **H ∥ ab** in $CuCrP_2(S_{1-X}Se_X)_6$ crystals at X=0, 0.02, and 0.05 on contrary to the case at X=0.1 and X=0.2. Compared to the length of antiferromagnetic profile in **H ∥ ab**, it is demonstrated that the length of antiferromagnetic path **H ⊥ ab** decreases gradually with increasing the selenium doping in $CuCrP_2(S_{1-X}Se_X)_6$. Notably, the mismatching in isotropic and anisotropic properties between antiferromagnetic and paramagnetic phases has also been reported for other $M_2P_2X_6$ single crystals [44]. According to **Figures 5(g)** and **5(h)**, the temperatures at maximum magnetic susceptibility values ($T_{Max}$) in case of **H ⊥ ab** measurements differ from those for **H ∥ ab** measurements as recorded in **Table 2**. It is elucidated that $T_{Max}$ for **H ∥ ab** is higher than that for **H ⊥ ab** except at X=0.05 where $T_{Max}$ values are the same in both situations. Besides, we can easily observe that $T_{Max}$ in both situations (**H ⊥ ab** and **H ∥ ab**) decreases with the increase in the selenium doping of $CuCrP_2(S_{1-X}Se_X)_6$ crystals.

**Table 2:** Néel temperature ($T_N$) values and the temperature values at maximum magnetic susceptibilities ($T_{Max}$) in both $\mathbf{H \perp ab}$ and $\mathbf{H \parallel ab}$ for $CuCrP_2(S_{1-X}Se_X)_6$.

| $X_{Se}$ in $CuCrP_2(S_{1-X}Se_X)_6$ | $T_N$ (K) | $T_{Max}$ (K) ($\mathbf{H \perp ab}$) | $T_{Max}$ (K) ($\mathbf{H \parallel ab}$) |
|---|---|---|---|
| 0 | 35.29 | 37.89 | 38.20 |
| 0.02 | 33.31 | 37 | 37.59 |
| 0.05 | 30.57 | 37 | 37 |
| 0.1 | 32.54 | 32.54 | 33.17 |
| 0.2 | 31.8 | 31.94 | 32.01 |

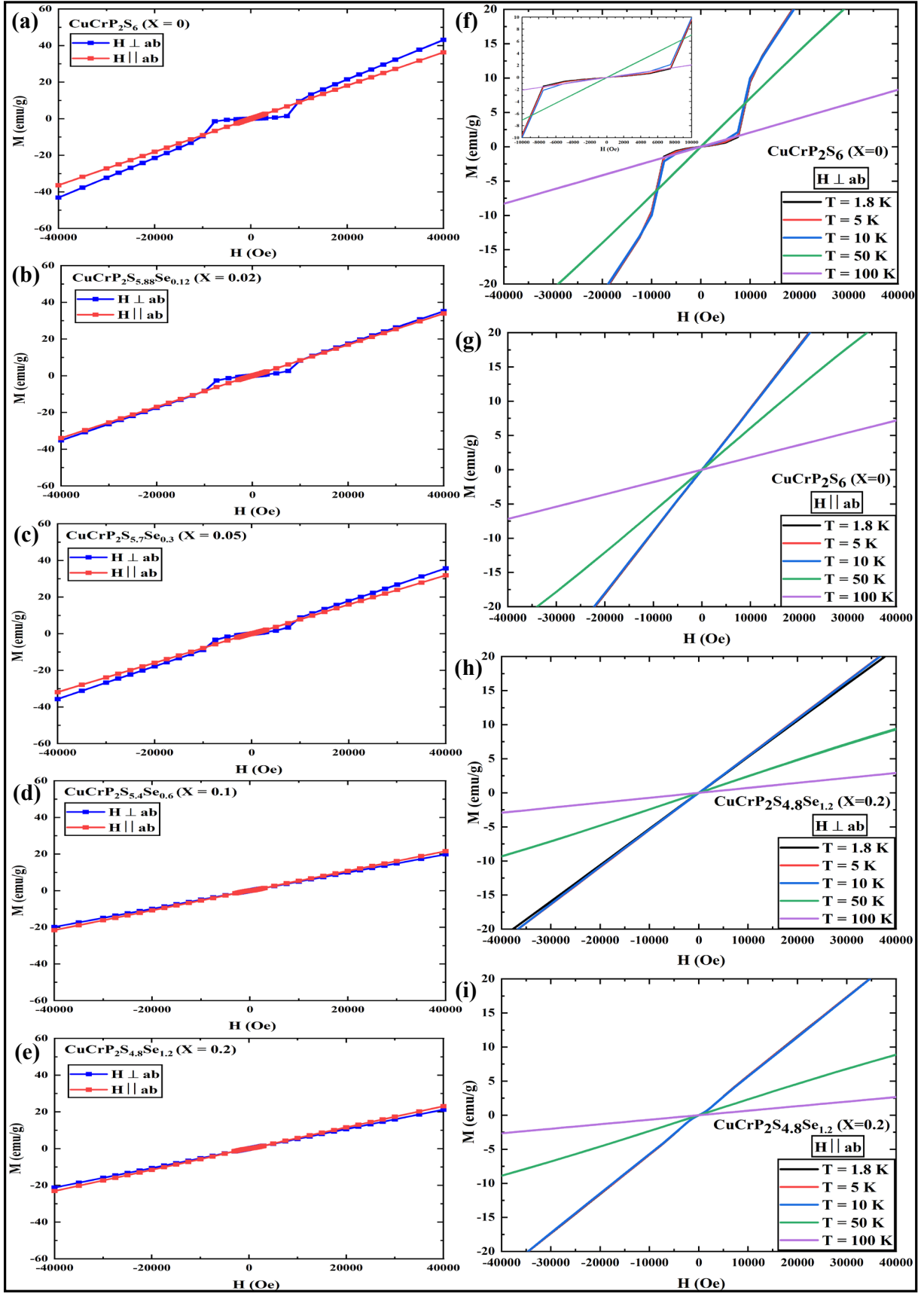

**Figure 6: (a)**, **(b)**, **(c)**, **(d)**, and **(e)** represent the hysteresis loops (**M** vs. **H**) at T=1.8 K of CuCrP$_2$(S$_{1-X}$Se$_X$)$_6$ crystals at X=0, 0.02, 0.05, 0.1, and 0.2, respectively, in both **H ⊥ ab** and **H ∥ ab**. **(f)** and **(g)** are **M** vs. **H** curves of CuCrP$_2$S$_6$ (X=0) for **H ⊥ ab** and **H ∥ ab**, respectively, at different temperatures (T=1.8 K, 5 K, 10 K, 50 K, and 100 K). **(h)** and **(i)** indicate **M** vs. **H** curves of Se-doped CuCrP$_2$S$_6$ (CuCrP$_2$S$_{4.8}$Se$_{1.2}$ (X=0.2)) for **H ⊥ ab** and **H ∥ ab**, respectively, at different temperatures (T=1.8 K, 5 K, 10 K, 50 K, and 100 K).

The relation between the magnetization (**M**) and the applied magnetic field (**H**) for CuCrP$_2$(S$_{1-X}$Se$_X$)$_6$ compounds has been demonstrated at T=1.8 K in **Figures 6((a)-(e))**. It is observed that a linear relation between **M** and **H** remains dominant in case of **H ∥ ab** for all selenium doping amounts at X=0 (**Figure 6(a)**), X=0.02 (**Figure 6(b)**), X=0.05 (**Figure 6(c)**), X=0.1 (**Figure 6(d)**), and X=0.2 (**Figure 6(e)**). On the other hand, in case of **H ⊥ ab**, clear transitions of the spin flop appear for the parent compound CuCrP$_2$S$_6$ as shown in **Figure 6(a)** as well as for the doped compounds at X=0.02 (**Figure 6(b)**) and X=0.05 (**Figure 6(c)**) while these transitions are overlapped and disappeared for the doped compounds at X=0.1 (**Figure 6(d)**) and X=0.2 (**Figure 6(e)**). Moreover, from **Figure 6(a)** to **Figure 6(e)**, with increasing the selenium doping amounts, the spin flop is gradually suppressed and becomes more broadened and very weak. Similar phenomenon has been reported for (Mn$_{1-X}$Ni$_X$)$_2$P$_2$S$_6$ single crystals [21]. The occurred change in the anisotropic properties is essentially detected due to the interplay between the schemes of the crystal field effect and the anisotropy of the single-ion together with the spin-orbit coupling effects (SOC) [44]. Also, in other van der Waals crystals, such as transition-metal mixed halides [45] and Cu$_{2x}$Fe$_{1-x}$PS$_3$ [46], the magnetic anisotropic properties were observed and tuned based on the changes in the compositional elemental contents. Furthermore, the temperature-dependence evolution of the relation between **M** and **H** has been illustrated in **Figure 6(f)** to **Figure 6(i)**. **Figure 6(f)** shows that the spin flop transitions of the parent compound CuCrP$_2$S$_6$ in case of **H ⊥ ab** appear only at T=1.8 K, T=5 K, and T=10 K while they disappear at T=50 K and T=100 K. Also, as the temperature increases, the spin flop is slightly suppressed until disappearing completely at temperatures higher than Néel temperature as presented in the inset of **Figure 6(f)**. On the other hand, no spin flop transitions appeared in case of **H ∥ ab (Figure 6(g))** for the pristine CuCrP$_2$S$_6$ as well as for the doped compound CuCrP$_2$S$_{4.8}$Se$_{1.2}$ in both **H ⊥ ab (Figure 6(h))** and **H ∥ ab (Figure 6(i))**. According to the linear profile between **M** and **H** for CuCrP$_2$S$_6$ (X=0) and CuCrP$_2$S$_{4.8}$Se$_{1.2}$ (X=0.2) for both **H ⊥ ab** and **H ∥ ab** states at different temperatures, the magnetic susceptibility is almost the same for the temperatures lower than Néel temperature (T=1.8 K, T=5 K, and T=10 K) while it is largely changed for the temperatures higher than Néel temperature (T=50 K and T=100 K). It is seen that with increasing the temperature, the magnetic susceptibility for CuCrP$_2$S$_6$ (X=0) and CuCrP$_2$S$_{4.8}$Se$_{1.2}$ (X=0.2) decreases independently on the crystallographic directions. In addition, the magnetic susceptibility of the pure CuCrP$_2$S$_6$ (X=0) structures is higher than that for the doped CuCrP$_2$S$_{4.8}$Se$_{1.2}$ (X=0.2) structure. By comparing the temperature variation effects with the effects of the doping amount change on the spin flop transitions, it is found that the change in the selenium doping amount can suppress these transitions larger than the suppression resulting from the change in the temperature. This confirms that the variations in selenium dopants are more efficient in modulating the spin flop transition than the temperature variations.

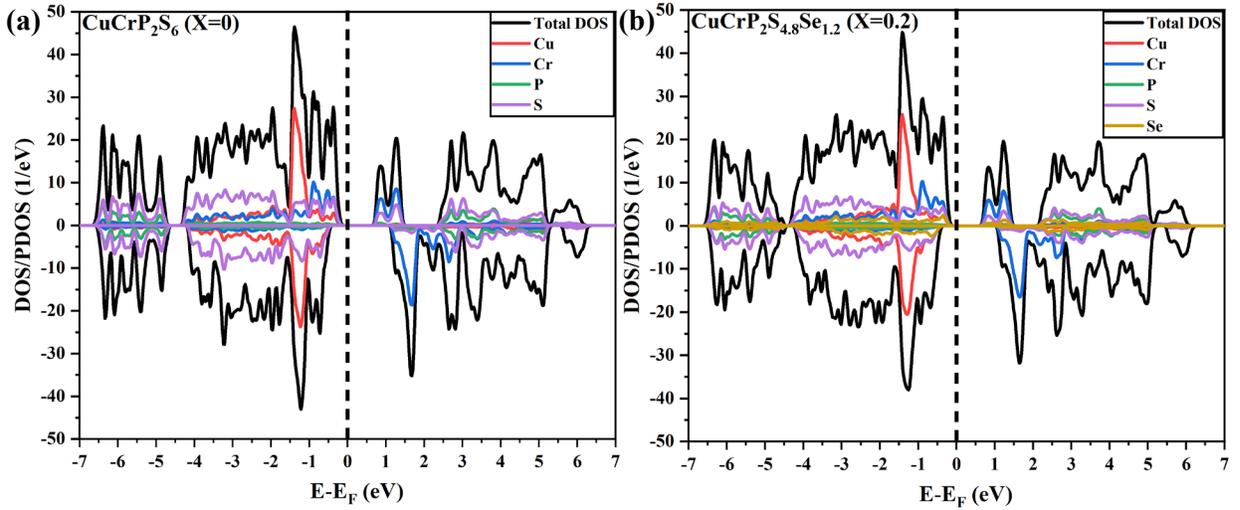

**Figure 7:** Total and projected density of states of **(a)** CuCrP$_2$S$_6$ and **(b)** Se-doped CuCrP$_2$S$_6$ (CuCrP$_2$S$_{4.8}$Se$_{1.2}$)

Total density of states (TDOS) and projected density of states (PDOS) of each element in **Figure 7** show the asymmetric characteristics between spin-up and spin-down states in both parent (**Figure 7(a)**) and doped (**Figure 7(b)**) systems. As shown in **Figure 7(a)**, the contribution of S is dominant at - 0.16 eV (top of the valence band) while the contribution of Cr becomes more dominant after Se doping at the same energy value as shown in **Figure 7(b)**. In addition, Cr has the pronounced contribution at the bottom of the conduction bands for the pristine and doped compounds at 0.67 eV and 0.66 eV, respectively. On the left-hand side (LHS) of **Figures 7(a)** and **7(b)**, the hybridization of spin-up states of Cr is the strongest, ranging from -0.46 eV to -0.93 eV for the undoped crystal and from -0.33 eV to -0.99eV for the doped one. On the other hand, the hybridization of spin-up and spin-down states of Cr in the right-hand side (RHS) becomes very strong in the energy range from 0.75 eV to 2.34 eV for the pristine structure and from 0.7 eV to 2.1 eV for the doped one. Also, the contribution of Cu is the largest and more efficient only during the energy ranges from -1.1 eV to -1.5 eV for CuCrP$_2$S$_6$ (**Figure 7(a)**) and from -1 eV to -1.53 eV for CuCrP$_2$S$_{4.8}$Se$_{1.2}$ (**Figure 7(b)**). Besides, in doped compound (**Figure 7(b)**), the contribution of P's spin-up states is only observed in its dominant form in the energy interval ranged from 3.6 eV to 3.8 eV while S's contribution becomes evidently dominant at the same energy interval in the undoped compound (**Figure 7(a)**). The S's hybridization has a great effect compared to other hybridizations in both parent (from -1.6 eV to -6.5 eV in LHS and from 2.95 eV to 6.2 eV in RHS) and doped (-1.9 eV to -6.5eV in LHS and 2.8 eV to 3.5 eV in RHS) structures. Moreover, the hybridization intensity of S contributions in LHS of **Figures 7((a) and (b))** is higher than that in RHS for both undoped and doped systems. Notably, the hybridization intensity of Se contributions is weaker than all hybridizations in the doped compound. In general, the hybridization intensities in parent structure decrease largely under Se addition due to strong orbital interactions between Se atoms and its neighbors. Finally, it is confirmed that the Se doping can greatly modify the asymmetric behavior of spin-up and spin-down states which has an effective role in the anisotropy in the pristine CuCrP$_2$S$_6$ crystals.

## IV. Conclusions

In summary, chemical vapor transport technique was used to synthesize high quality crystals of $CuCrP_2(S_{1-X}Se_X)_6$ (where X=0, 0.02, 0.05, 0.1, and 0.2). These crystals were characterized by SEM, EDS, XRD, HRTEM, SAED and XPS. The magnetic measurements were carried out in perpendicular and parallel directions between the external magnetic field (**H**) and the crystallographic plane (**ab**). In addition, spin polarized DFT calculations were performed to study the effect of selenium substitution on the magnetic properties of the pure $CuCrP_2S_6$. The experimental results revealed that magnetic anisotropy can be greatly controlled, especially in the antiferromagnetic phase, via selenium doping. Besides, $T_N$ values decrease with increasing the selenium concentrations. $T_{Max}$ for **H ∥ ab** is higher than that for **H ⊥ ab** except at X=0.05 where they have the same values. In both **H ∥ ab** and **H ⊥ ab** cases, $T_{Max}$ values decrease with increasing the selenium concentrations. The suppressions of spin flop transitions, which appear only in case of **H ⊥ ab**, can be tunned predominantly due to the increase in selenium concentrations and slightly due to the increase in the temperature. The transitions of spin flop appeared only at $T<T_N$ (antiferromagnetic phase) while they disappeared at $T>T_N$ (paramagnetic phase). Also, the appearance of these transitions occurred only at X=0, 0.02, 0.05 while their disappearance was observed at X = 0.1, 0.2. DFT calculations confirmed that selenium doping in $CuCrP_2S_6$ has a high contribution to change the asymmetric characteristics between spin up and spin down states. It is evident that the tunable magnetic anisotropy in this material makes it a suitable candidate for spintronic devices applications.

## V. Author Contributions

I. S. E. set up the experimental design, material synthesis, data analysis, computational calculations and writing the original draft of the manuscript; L. S. contributed in conceptualization, supervision and manuscript review & editing.

## VI. Acknowledgements

We would like to express our thanks to Dr. Zia ur Rehman (Visiting Researcher at National Synchrotron Radiation Laboratory, University of Science and Technology of China, Hefei, Anhui 230029, China) and Dr. Ahmed Shahboub (Postdoctoral Researcher at Hefei Institutes of Physical Sciences, Chinese Academy of Sciences, Hefei, China) for their fruitful discussions.